\documentstyle[12pt]{article}
\begin{document}

\title{Kaon versus Antikaon Production at SIS Energies\footnote{supported
by GSI Darmstadt}}
\author{W. Cassing, E. L. Bratkovskaya, U. Mosel, S. Teis and A. Sibirtsev\\
Institut f\"{u}r Theoretische Physik, Universit\"{a}t Giessen\\
35392 Giessen, Germany}
\date{}

\maketitle

\begin{abstract}
We analyse the production and propagation of kaons and antikaons in
Ni~+~Ni reactions from 0.8--1.85 GeV/u within a coupled channel
transport approach including the channels $BB \to K^+YN, \ \pi B\to
K^+Y, \ BB \to NN K \bar{K}, \ \pi B\to N K\bar{K}, \ K^+B\to K^+B, \
\bar{K} B\to \bar{K}B, \ Y N\to \bar{K} NN, \ \pi \pi\to K \bar{K}$
as well as $\pi Y\to \bar{K}N$ and $\bar{K} N\to \pi Y$ for the antikaon
absorption.  Whereas the experimental $K^+$ spectra can be reproduced
without introducing any selfenergies for the mesons in Ni~+~Ni
collisions from 0.8 to 1.8 GeV/u, the $K^-$ yield is underestimated by
a factor of 5--7 at 1.66 and 1.85 GeV/u.  However, introducing density
dependent antikaon masses as proposed by Kaplan and Nelson, the
antikaon spectra can be reasonably well described.
\end{abstract}
\vspace{1cm}

\newpage
\section{Introduction}
The study of hot and dense nuclear matter via relativistic
nucleus-nucleus collisions is the major aim of high energy heavy-ion
physics. Nowadays, the search for a restoration of chiral symmetry at
high baryon density or for a phase transition to the quark-gluon plasma
(QGP) is of specific interest. However, any conclusions about the
hadron properties at high temperature or baryon densities must rely on
the comparison of experimental data with theoretical approaches based
on nonequilibrium kinetic theory. Among these, the covariant RBUU
approach~\cite{Cass,lang,Teis,Ko1,Ko2,Ko3,Weber1,Weber2,Tomo}, the
QMD~\cite{Aich} or RQMD model~\cite{RQMD}, and more recently the HSD
approach~\cite{Ehehalt} have been successfully used in the past. As a
genuine feature of transport theories there are two essential
ingredients: i.e.  the  baryon (and meson) scalar and vector
selfenergies -- which are neglected in many approaches -- as well as
in-medium elastic and inelastic cross sections for all hadrons
involved.

Selfenergy effects in the production of particles have been found
previously for antiprotons by a number of
groups~\cite{Teis,Ko3,cassinga,Batko95} though the actual magnitude of
the attractive $\bar{p}$-potential in the nuclear medium is still a
matter of debate.  The first dynamical studies on antikaon production
in nucleus-nucleus collisions have been performed since more than a
decade ago without including any selfenergies for the mesons
produced~\cite{old1,old2,huang92}.  However, as advocated
in~\cite{GB1,Brown,Kaplan,Brown1,lee,Ko,mrho,lutz} also antikaons should
feel strong attractive forces in the medium so that their production
threshold should be reduced at finite baryon density. A first
exploratory study with respect to antikaon selfenergies has been
performed by Li, Ko and Fang in Ref.~\cite{Kolix} with the result that
sizeable $K^-$ potentials are needed to explain the experimental
spectra from~\cite{Schro} for Ni~+~Ni at 1.85 GeV/u. Similar, but less
pronounced results have been obtained in Ref.~\cite{Ehehalt} at AGS
energies where especially the kaon/pion ratio is much better reproduced
when including kaon selfenergies. However, at 10--15 GeV/u there is
also a sizeable kaon production from meson-meson channels with partly
unknown cross sections, which contribute to the $K^+K^-$ yield. Thus
heavy-ion collisions in the range of only 1--2 GeV/u are more
favourable systems for the study of kaon dynamics since the number of
production channels is much more restricted while already considerable
baryon densities (up to 3 $\times \rho_0 \approx 0.5 fm^{-3}$) can be
reached in the compression phase~\cite{Cass}.

The most serious problem related to $K^-$ production at 'subthreshold'
energies are the baryon-baryon and pion-baryon elementary production
cross sections close to threshold where no experimental data are
available so far. It has recently been argued that extrapolations from
high energy data~\cite{zwermann} used in~\cite{Kolix} overestimate the
elementary $K^-$ yield by more than an order of magnitude close to
threshold~\cite{Sibirt1,Sibirt2,Sibirt4} such that a reanalysis of this
question in nucleus-nucleus reactions appears necessary.  In previous
works we have calculated -- within boson exchange models -- the near
threshold production cross section for antiprotons~\cite{Lyk96}, vector
mesons ($\rho, \omega, \Phi$)~\cite{Sibirt3} as well as
antikaons~\cite{Sibirt4} from nucleon-nucleon as well as pion-nucleon
reactions (cf. also Ref.~\cite{texas} in case of $K^+$ mesons). Since
within the boson exchange model one can interpolate between different
isospin channels and thus compare to a much larger set of experimental
data, the results of these studies should be more reliable than the
early parametrizations~\cite{zwermann,randrup,schur}, that partly are
in severe contradiction to the available phase space.

Our work thus is organized as follows: we start in Sect. 2 with a brief
description of the transport approach employed and specify the kaon
selfenergies incorporated in the calculation as well as the various
kaon production and reabsorption channels that are taken into account.
Section 3 is devoted to a presentation of our calculated results for
Ni~+~Ni at 0.8--1.85 GeV/u in comparison to published experimental
data. We will include also a differential analysis of the various
production channels contributing to the antikaon production and explore
the effect of kaon selfenergies.  Section 4, finally, is devoted to a
summary and discussion of open problems.

\section{Ingredients of the transport approach}

In this paper we perform our analysis along the line of the
HSD\footnote{Hadron String Dynamics} approach~\cite{Ehehalt} which is
based on a coupled set of covariant transport equations for the
phase-space distributions $f_{h} (x,p)$ of hadron
$h$~\cite{Weber1,Ehehalt}, i.e.
\begin{eqnarray}  \label{g24}
\lefteqn{\left\{ \left( \Pi_{\mu}-\Pi_{\nu}\partial_{\mu}^p U_{h}^{\nu}
-M_{h}^*\partial^p_{\mu} U_{h}^{S} \right)\partial_x^{\mu}
+ \left( \Pi_{\nu} \partial^x_{\mu} U^{\nu}_{h}+
M^*_{h} \partial^x_{\mu}U^{S}_{h}\right) \partial^{\mu}_p
\right\} f_{h}(x,p) } \nonumber \\
&& = \sum_{h_2 h_3 h_4\ldots} \int d2 d3 d4 \ldots
 [G^{\dagger}G]_{12\to 34\ldots}
\delta^4(\Pi +\Pi_2-\Pi_3-\Pi_4 \ldots )  \nonumber\\
&& \times \left\{ f_{h_3}(x,p_3)f_{h_4}(x,p_4)\bar{f}_{h}(x,p)
\bar{f}_{h_2}(x,p_2)\right.  \nonumber\\
&& -\left. f_{h}(x,p)f_{h_2}(x,p_2)\bar{f}_{h_3}(x,p_3)
\bar{f}_{h_4}(x,p_4) \right\} \ldots\ \ .
\end{eqnarray}
In Eq.~(\ref{g24}) $U_{h}^{S}(x,p)$ and $U_{h}^{\mu}(x,p)$ denote the
real part of the scalar and vector hadron selfenergies, respectively,
while $[G^+G]_{12\to 34\ldots} \delta^4 (\Pi
+\Pi_2-\Pi_3-\Pi_4\ldots )$ is the 'transition rate' for the process
$1+2\to 3+4+\ldots$ which is taken to be on-shell in the
semiclassical limit adopted. The hadron quasi-particle properties in
(\ref{g24}) are defined via the mass-shell constraint~\cite{Weber1},
\begin{equation}   \label{g25}
\delta (\Pi_{\mu}\Pi^{\mu}-M_{h}^{*2} ) \ \ ,
\end{equation}
with effective masses and momenta (in local Thomas-Fermi approximation)
given by
\begin{eqnarray}\label{g26}
M_{h}^* (x,p)&=&M_h + U_h^{{S}}(x,p) \nonumber \\
\Pi^{\mu} (x,p)&=&p^{\mu}-U^{\mu}_h (x,p)\ \ ,
\end{eqnarray}
while the phase-space factors
\begin{equation}
\bar{f}_{h} (x,p)=1 \pm f_{{h}} (x,p)
\end{equation}
are responsible for fermion Pauli-blocking or Bose enhancement,
respectively, depending on the type of hadron in the final/initial
channel. The dots in Eq.~(\ref{g24}) stand for further contributions to
the collision term with more than two hadrons in the final/initial
channels. The transport approach (\ref{g24}) is fully specified by
$U_{h}^{S}(x,p)$ and $U_{h}^{\mu}(x,p)$ $(\mu =0,1,2,3)$, which
determine the mean-field propagation of the hadrons, and by the
transition rates $G^\dagger G\,\delta^4 (\ldots )$ in the collision
term, that describes the scattering and hadron production/absorption
rates.

The scalar and vector mean fields $U_{h}^{S}$ and $U^\mu_{h}$ for
baryons are taken from Ref.~\cite{Ehehalt} and don't have to be
specified here again.  In the present approach we propagate explicitly
pions, kaons, and $\eta$'s and assume that the pions as Goldstone
bosons do not change their properties in the medium; we also discard
selfenergies for the $\eta$-mesons in the present calculation which
have a minor effect on the kaon dynamics.  The kaon-baryon interactions
are described along the line of Kaplan and Nelson~\cite{Kaplan} and
have to be specified more explicitly.

\subsection{Kaon selfenergies}

As in case of antiprotons there are a couple of models for the kaon and
antikaon selfenergies~\cite{GB1,Brown,Kaplan,Brown1,lee,Ko,mrho,lutz},
which differ in the actual magnitude of the selfenergies, however,
agree on the relative signs for kaons and antikaons. Thus in line with
the kaon-nucleon scattering amplitude the $K^+$ potential should be
slightly repulsive at finite baryon density whereas the antikaon should
see an attractive potential in the nuclear medium. Without going into a
detailed discussion of the various models we adopt the more practical
point of view, that the actual $K^+$ and $K^-$ selfenergies are unknown
and as a guide for our analysis adopt the approach by Kaplan and
Nelson~\cite{Kaplan}.

Starting from a $SU(3)_{L}\times SU(3)_{R}$ chiral Lagrangian and using
chiral perturbation theory Kaplan and Nelson~\cite{Kaplan} have derived
an effective meson-baryon Lagrangian up to first order in the baryon
scalar density, which they claim to be valid up to $\sim 7\rho_0$.
Since the coefficients in this Lagrangian are approximately known
experimentally (within an uncertainty of about 30\%), one can model a
Lagrangian of lower complexity, but with the same properties on the
mean-field level. Such limits lead to the following dispersion relation
for kaons in the nuclear medium~\cite{Nelson,Liko}:
\begin{eqnarray}\label{mg1}
\omega_{{K}^+}(x,{\bf p})&=& \left\{{\bf p}^2 + m_{K}^2\left(
1-{\Sigma_{K N}\over f^2_{K}
m_{K}^2}\rho_{S}(x) + \left({3\rho_{B}(x)\over 8
f_{K}^2m_{K}}\right)^2\right)\right\}^{1/2}+ {3\over
8}{\rho_{B}(x)\over f^2_{K}}\ \ ,\nonumber\\
\omega_{{K}^-}(x,{\bf p})&=& \left\{{\bf p}^2 + m_{K}^2\left(
1-{\Sigma_{K N}\over f^2_{K}
m_{K}^2}\rho_{S}(x) + \left({3\rho_{B}(x)\over 8
f_{K}^2m_{K}}\right)^2\right)\right\}^{1/2}- {3\over 8}
{\rho_{B}(x)\over f^2_{K}}\ \ ,
\end{eqnarray}
with $m_{K}$ denoting the bare kaon mass, $f_{K}\approx$ 93 MeV and
$\Sigma_{K N}\approx $ 350 MeV, while $\rho_{S}(x)$ and $\rho_{B}(x)$
are the scalar and vector baryon densities at space-time position $x$,
respectively. We note, that the kaon quasi-particle dynamics in the
transport approach (1) are completely determined by the space and
momentum derivatives of (\ref{mg1}) according to the Hamilton
equations.

In order to evaluate the dispersion relations (\ref{mg1}) we need a
model to compute the scalar density $\rho_S$ as a function of the
baryon density $\rho_B$. Here, the relativistic RBUU
model~\cite{Weber1} provides a useful guide. Within a good
approximation the nucleon effective mass for nuclear matter at
temperature T = 0 can be written as
\begin{equation}
m^*_N = {m_N \over 1 + 0.52 \displaystyle{\rho_B\over \rho_0}},
\label{mnstar}
\end{equation}
where $m_N$ is the vacuum nucleon mass and $\rho_0 \approx 0.16 \ fm^{-3}$.
The scalar density then can be evaluated by
\begin{equation}
\rho_S(p_F) = \frac{g}{2 \pi^2} \int_0^{p_f} \ dp \ p^2
\frac{m^*}{\sqrt{p^2+m^{*2}}},
\label{rhoS}
\end{equation}
with the Fermi momentum
\begin{equation}
p_F = \left({3\over 2 \pi^2} \ \rho_B\right)^{1/3}
\label{pfermi}
\end{equation}
and $g = 4$ for isospin symmetric matter.

In the restframe of the kaons the $K^+$ and $K^-$ energy/mass changes
almost linearly with the baryon density as shown in Fig.~\ref{Fig1}.
Since the $K^+$ mass changes only very modestly with the baryon
density, we will discard any selfenergy effects on the $K^+$ mesons in
the present analysis; this approximation will be confronted with the
experimental data in Sect.~3.

For practical purposes, furthermore, we include the $K^-$ selfenergy
in a density dependent quasi-particle mass as
\begin{equation}
\label{kmass}
m^*_K(\rho_B) = m_K^0 \left(1 - \alpha \frac{\rho_B}{\rho_0}\right)
\end{equation}
with $\alpha \approx $ 0.2 (cf. lower dashed line in Fig.~\ref{Fig1})
\footnote{In Ref.~\cite{Ehehalt} we have used $\alpha$ = 0.16 in order
to get a linear fit up to 6$\times \rho_0$ as needed for the AGS energy
regime.}. Linear fits to the antikaon selfenergies from~\cite{lutz,mrho}
lead to different values for the parameter $\alpha$ in the range 0.1
$\leq \alpha \leq 0.4$. We note that the dropping of the antikaon mass
is associated with a corresponding scalar energy density in the
baryon/meson Lagrangian, such that the total energy-momentum is
conserved during the heavy-ion collision (cf.~\cite{Ehehalt}).

\subsection{Kaon and antikaon reaction channels}

First, the individual production channels of the kaons have to be
specified. Here, we prefer to express the cross sections as a function
of the scaled invariant energy squared $s_0/s$, since the change of the
quasi-particle mass then can be incorporated in the threshold energy
$\sqrt{s_0}$ for the particular channel. This recipe might be still a
matter of debate; however, our findings in
Refs.~\cite{Sibirt1,Sibirt2,Sibirt4,Lyk96,Sibirt3} indicate, that the
production is essentially dominated by phase space close to threshold
and thus a scaling in $s_0/s$ should be a good approximation.

The isospin averaged  production cross section of a $K^+ \Lambda$ and
$K^+ \Sigma$ pair in a nucleon-nucleon collision is related to the
measured isospin channels as:
\begin{eqnarray}
&&\sigma_{NN \to K^+ \Lambda N} = {3\over 2} \sigma_{pp \to K^+ \Lambda p}
 \label{nnkp}\\
&&\sigma_{NN \to K^+ \Sigma N} = {3\over 2} \left(
\sigma_{pp \to K^0 \Sigma^+ p} + \sigma_{pp \to K^+ \Sigma^0 p} \right).
 \label{nnkp0}
\end{eqnarray}
Following~\cite{Sibirt2} the reaction cross section can be
approximated by
\begin{eqnarray}
&& \sigma_{pp \to K^+ \Lambda p}(s) = 732 \
{\left( 1 - \frac {s_{01}} {s} \right) }^{1.8}
{\left(\frac {s_{01}} {s} \right) }^{1.5} \ \ [{\mu}b]
  \label{Sppkp1}\\
&& \sigma_{pp \to K^0 \Sigma^+ p}(s) = 338.46 \
{\left( 1 - \frac {s_{02}} {s} \right) }^{2.25}
{\left(\frac {s_{02}} {s} \right) }^{1.35} \ \ [{\mu}b]
  \label{Sppkp2}\\
&& \sigma_{pp \to  K^+ \Sigma^0 p}(s) = 275.27 \
{\left( 1 - \frac {s_{02}} {s} \right) }^{1.98}
{\left(\frac {s_{02}} {s} \right) } \ \ [{\mu}b]
  \label{Sppkp3}\\
\end{eqnarray}
with $\sqrt{s_{01}} = m_\Lambda - m_N + m^0_K$ and $\sqrt{s_{02}} =
m_\Sigma + m_N + m^0_K$. According to isospin relations the $N\Delta$
and $\Delta\Delta$ production channels get additional factors of 3/4
and 1/2, respectively. The elementary cross sections for the pion
induced channels $\pi N \to K^+ Y$ have been computed by Tsushima et
al. in Ref.~\cite{Ts1}, which  we adopt for our present study.

The isospin averaged $K^-$ production cross section from
nucleon-nucleon collisions is taken from~\cite{Sibirt4} in the
parametrization ($K = (K^0, K^+), \bar{K} = (\bar{K}^0, K^-)$)
\begin{equation}
\label{kmin}
\sigma_{NN \to NN K\bar{K}} (s) \approx a \left(1 - \frac{s_0}{s}
\right)^{3.17} \left(\frac{s_0}{s}\right)^{1.96}
\end{equation}
with $a$ = 1.5 mb and $\sqrt{s_0} = 2 m_N + m^0_K + m^*_K$. The
channels $N \Delta$ and $\Delta \Delta$ are taken to be the same as
(\ref{kmin}) due to the $K\bar{K}$ pair in the final state. The latter
channels play an essential role due to the formation of resonance
matter in the heavy-ion collision zone~\cite{Eheh93} and due to lack of
experimental information provide a major source of uncertainty in the
present analysis. We will come back to this question in Sec.~3.

Our parametrizations for the $K^+$ and $K^-$ cross section from $pp$
collisions are displayed in Fig.~\ref{Fig2} by the solid lines in
comparison to the inclusive experimental data for $K^+$ production
(open squares) and $K^-$ production (full circles) as a function of the
invariant energy above threshold.  The cross section for $K ^+$
production includes both $\Lambda $ and $\Sigma $-hyperon reaction
channels. The dash-dotted line (denoted by Z\&S) reflects the
parametrization from Ref.~\cite{zwermann} that has been used so
far~\cite{Kolix,old2} for studies of $K^-$ production in heavy-ion
collisions and close to threshold is much larger than our cross section
for $K^-$ by more than an order of magnitude. Our novel parametrization
for the inclusive $K^-$ cross section is essentially based on the
results of the boson-exchange model from~\cite{Sibirt4} for the isospin
averaged cross section close to threshold (open square denoted by OBE
in Fig.~\ref{Fig2}).  In this OBE-model the different exclusive isospin
channels are related to each other via the same Feynman graphs and a
comparison to a much larger set of experimental data can be
established. As an example for antikaon production we show in the left
upper part of Fig.~\ref{Fig3} the result of the calculation for the
reaction $pp \to pn K^+ \bar{K}^0$ in comparison to the experimental
data from~\cite{Giacom}. Also other channels like $pp \to p \Lambda
K^+, pp \to p \Sigma^0 K^+$ or $pp \to  K^0 \Sigma^+ p$ are described
reasonably well within this OBE approach as seen from Fig.~\ref{Fig3}.
We thus expect our new parametrization to be more realistic than that
of Ref.~\cite{zwermann}.  However, a detailed experimental study close
to threshold energies should be performed to obtain final numbers for
the elementary production channels. We finally note that the ratio of
the $K^+/K^-$ production cross section increases when going closer to
threshold; this is expected because the final phase space for $K^+$
production is of three-body type while it is a four-body phase space in
case of antikaon production.

The experimental $\pi^-p\to pK^0K^-$ cross section, furthermore, can be
expressed by~\cite{Sibirt4},
\begin{equation}
\label{par0}
\sigma ({\pi^-} p\to pK^0K^-) = 1.121 \
{\left( 1- \frac {s_0} {s} \right)}^{1.86}
{\left( \frac {s_0} {s} \right)}^2  \ \ [mb],
\end{equation}
where $\sqrt{s}$ is the invariant mass of the $\pi N$ system and
$\sqrt{s_0}=m_N+m_K^0 + m^*_K$. Exploring isospin
symmetries~\cite{Sibirt4} the other cross sections can be related to
$\sigma(\pi^- p \to p K^0 K^-)$ by:
\begin{eqnarray}
\label{iso}
 2 \sigma ({\pi}^+ p \to p K^+ {\bar K^0}) = &
2 \sigma ({\pi}^+ n \to n K^+ {\bar K^0}) =  &
\sigma ({\pi}^+ n \to p K^+ K^-) =  \nonumber  \\
 \sigma ({\pi}^+ n \to p K^0 {\bar K^0}) = &
\sigma ({\pi}^0 p \to n K^+ {\bar K^0}) =  &
4 \sigma ({\pi}^0 p \to p K^+ K^- ) = \nonumber  \\
 4 \sigma ({\pi}^0 p \to p K^0 {\bar K^0} ) = &
\sigma ({\pi}^0 n \to p K^0  K^-) =           &
4 \sigma ({\pi}^0 n \to n K^+ K^- ) =  \nonumber \\
 4 \sigma ({\pi}^0 n \to n  K^0 {\bar K^0} ) = &
2 \sigma ({\pi}^- p \to p K^0 K^- )  =         &
\sigma ({\pi}^- p \to n K^+ K^- ) =   \nonumber \\
 \sigma ({\pi}^- p \to n K^0 {\bar K^0} ) =    &
2 \sigma ({\pi}^- n \to n K^0  K^-).           &
\end{eqnarray}
These isospin relations have been found in~\cite{Sibirt4} to be well in
line with the experimental data from~\cite{landolt}.

A further important production channel is given by the flavor exchange
reaction $\pi \Lambda \to K^- N$ and $\pi \Sigma \to K^-N$, where the
strange quark from the hyperon is exchanged with a light (u,d) quark.
The inverse reaction is the dominant channel for $K^-$ absorption on
nucleons. The latter absorption cross section from Ref.~\cite{absorp}
is displayed in Fig.~\ref{Fig4} by the solid line together with the
elastic $K^-N$ cross section (dashed line) as a function of the
$K$-meson momentum with respect to the nucleon at rest. It should be
noted that these cross sections are rather well known
experimentally~\cite{landolt} and that our parametrization provides an
optimal fit through the data.

Apart from the antikaon final state interactions shown in
Fig.~\ref{Fig4}, $K^+$ elastic scattering with nucleons also has an
impact on the final kaon spectra. The elastic cross section employed is
displayed in Fig.~\ref{Fig4} by the dotted line and indicates that
$K^+$ rescattering is of minor importance for our present study;
however, it is explicitly included in the actual calculations.

By using detailed balance, i.e.
\begin{equation}
\sigma_{\pi Y \to \bar{K} N}(s) = \frac
{\left[ s-(m_K^0+m_N)^2 \right] \left[ s-(m_K^0-m_N)^2 \right] }
{\left[ s-(m_\pi+m_Y)^2) \right] \left[ s-(m_\pi-m_Y)^2) \right]}
\ \sigma_{\bar{K} N \to \pi Y}(s),
\label{spiY}
\end{equation}
the $\pi$-hyperon production channels can be computed from the
parametrizations of the isospin averaged $\bar{K} N \to \pi Y $ cross
sections~\cite{CugnL}.  The corresponding results are displayed in
Fig.~\ref{Fig5}  as a function of the invariant energy $\sqrt{s}$
(solid and dashed line).  Due to the rather well known data for $K^- N$
scattering~\cite{landolt} these flavor exchange reactions are expected
to be well determined in the actual calculations.

In addition, we include the $K\bar{K}$ production by $\pi \pi$
reactions; the isospin averaged cross section for this channel can be
parametrized by~\cite{GB1}
\begin{equation}
\label{pipi}
\bar{\sigma}_{\pi \pi \to K\bar{K}} \approx a \
\left(1- \frac{s_0}{s}\right)^{0.76}
\end{equation}
with $a$ = 2.7 mb and $s_0 = (m_K + m_{\bar{K}})^2$. This cross section
is shown also in Fig.~\ref{Fig5} by the dash-dotted line. As we will
find in Sect. 3, the contribution of the $\pi \pi$ channel will be less
than 5\% up to bombarding energies of 1.85 GeV/u and could be neglected here.

We, furthermore, have incorporated the antikaon production by
hyperon-baryon collisions, which is evaluated in a OBE approach, too.
The resulting cross sections are displayed in Fig.~\ref{Fig6} by the
dashed and solid line which maxima below 1 mb. We will find in Sect.~3
that also these channels are of minor importance since their
contribution is below 1\% up 1.85 GeV/u.

We note again that the cross sections shown in
Figs.~\ref{Fig2}--\ref{Fig6} should account reasonably well for the
vacuum scattering processes, but it is not yet clear if a scaling of
the cross sections with their threshold values $\sqrt{s_0}$, which
include the in-medium hadron masses, and the corresponding scaling of
the $\Delta N$ cross sections, is the right recipe to adopt.  This
holds for the antikaon production as well as reabsorption channels and
has to be kept in mind when comparing our calculations to experimental
data.

\section{$\pi^-, K^+$ and $K^-$ yields for Ni~+~Ni reactions}

The calculation of 'subthreshold' particle production is described in
detail in Ref.~\cite{Cass,Cass90} and has to be treated perturbatively
in the energy regime of interest here due to the small cross sections
involved. Since we work within the parallel ensemble algorithm, each
parallel run of the transport calculation can be considered
approximately as an individual reaction event, where binary reactions
in the entrance channel at given invariant energy $\sqrt{s}$ lead to
final states with 2 (e.g. $K^+ Y$ in $\pi B$ channels), 3 (e.g. for
$K^+ YN$ channels in $BB$ collisions) or 4 particles (e.g. $K\bar{K}NN$
in $BB$ collisions) with a relative weight $W_i$ for each event $i$ which
is defined by the ratio of the production cross section to the total
hadron-hadron cross section\footnote{The actual final states are
chosen by Monte Carlo according to the 2, 3, or 4-body phase space.}.
The perturbative treatment now implies that in case of strangeness
production channels the initial hadrons are not modified in the
respective final channel. On the other hand, each strange hadron is
represented by a testparticle with weight $W_i$ and propagated
according to the Hamilton equations of motion. Elastic and inelastic
reactions with pions, $\eta$'s or nonstrange baryons are computed in
the standard way~\cite{Cass,wolf} and the final cross section is
obtained by multiplying each testparticle with its weight $W_i$.  In
this way one achieves a realistic simulation of the strangeness
production, propagation and reabsorption during the heavy-ion
collision, where only the dynamical feedback of the strange hadrons to
the nonstrange mesons and baryons is neglected.

We start our analysis with the system Ni~+~Ni at 1.85 GeV/u without
including any selfenergies for the antikaons. The inclusive
Lorentz-invariant cross for negative pions and antikaons in the
nucleus-nucleus cms (for $\theta_{lab} = 0^o$) is shown in
Fig.~\ref{Fig7} by the solid lines in comparison to the data of
Ref.~\cite{Schro} that were taken at $0^o$ in the laboratory system and
have been transformed to the nucleus-nucleus cms. We note that our
calculations yield an anisotropic $\pi^-$ angular distribution in this
reference frame in line with the analysis in Ref.~\cite{Teis96};
however, the $K^-$ angular distribution is found to be isotropic within
the numerical accuracy.

Whereas the $\pi^-$ spectra in Fig.~\ref{Fig7} are reasonably well
reproduced -- as in many other reactions in this energy
regime~\cite{wolf,Ehehalt1,Teis96} -- the antikaon spectra are
underestimated by up to a factor of 6--7. This finding agrees
qualitatively with that of Li et al. in Ref.~\cite{Kolix}, who also
underestimated these antikaon data substantially when using a vacuum
$K^-$ mass, even when adopting the parametrization of the elementary
cross section from Ref.~\cite{zwermann} (dash-dotted line in
Fig.~\ref{Fig2}).

It is, however, interesting to have a look at the contributions from
the different production channels in this case (cf. Fig.~\ref{Fig8}) in
comparison to the experimental data from~\cite{Schro} (full squares)
and the preliminary data from the KaoS collaboration~\cite{Senger1} for
Ni~+~Ni at 1.8 GeV/u.  Here the $BB$ channels are approximately in the
same order of magnitude as the $\pi B$ channels, but the $\pi Y$
channels provide the dominant contribution as also found in
Ref.~\cite{old2}. This is due to the fact that in more central
collisions the pion density reaches about $0.15 fm^{-3}$ while the
hyperons have almost the same abundancy as the $K^+$ mesons.  Thus a
substantial amount of hyperons suffer a quark exchange ($s \to u,d$)
when propagating out of the nuclear medium. We note, furthermore, that
for Ni~+~Ni at 1.85 GeV/u the $\pi \pi$-antikaon production channel
contributes by about 3\% whereas the hyperon-nucleon production
channels are below 1\%.

In order to demonstrate the sensitivity of our results to the $BB$
production cross section employed we show in Fig.~\ref{Fig8} by the
upper dot-dot-dashed line (denoted by Z\&S)  the total $K^-$ spectrum
when adopting the parametrization from Ref.~\cite{zwermann}. Here the
$BB$ channel becomes the most important one whereas the pion-hyperon
channel is about 30\% on average. Even in this limit the experimental
spectra are underestimated by about a factor of 4.

In order to demonstrate more clearly the relevant range of the
elementary $K^-$ cross section above threshold, we display in
Fig.~\ref{Fig9} the distribution in the baryon-baryon collision number
versus invariant energy $\sqrt{s}$, i.e.  $dN/d\sqrt{s}$ (histogram),
together with the parametrization from Ref.~\cite{zwermann} (dot-dashed
line) and Eq.~(\ref{kmin}) (solid line). It is clearly seen that the
dominant contributions stem from collisions with invariant energies far
below the first experimental data point and that an ambiguity remains
as long as our parametrization is not controlled by experimental
measurements close to threshold.

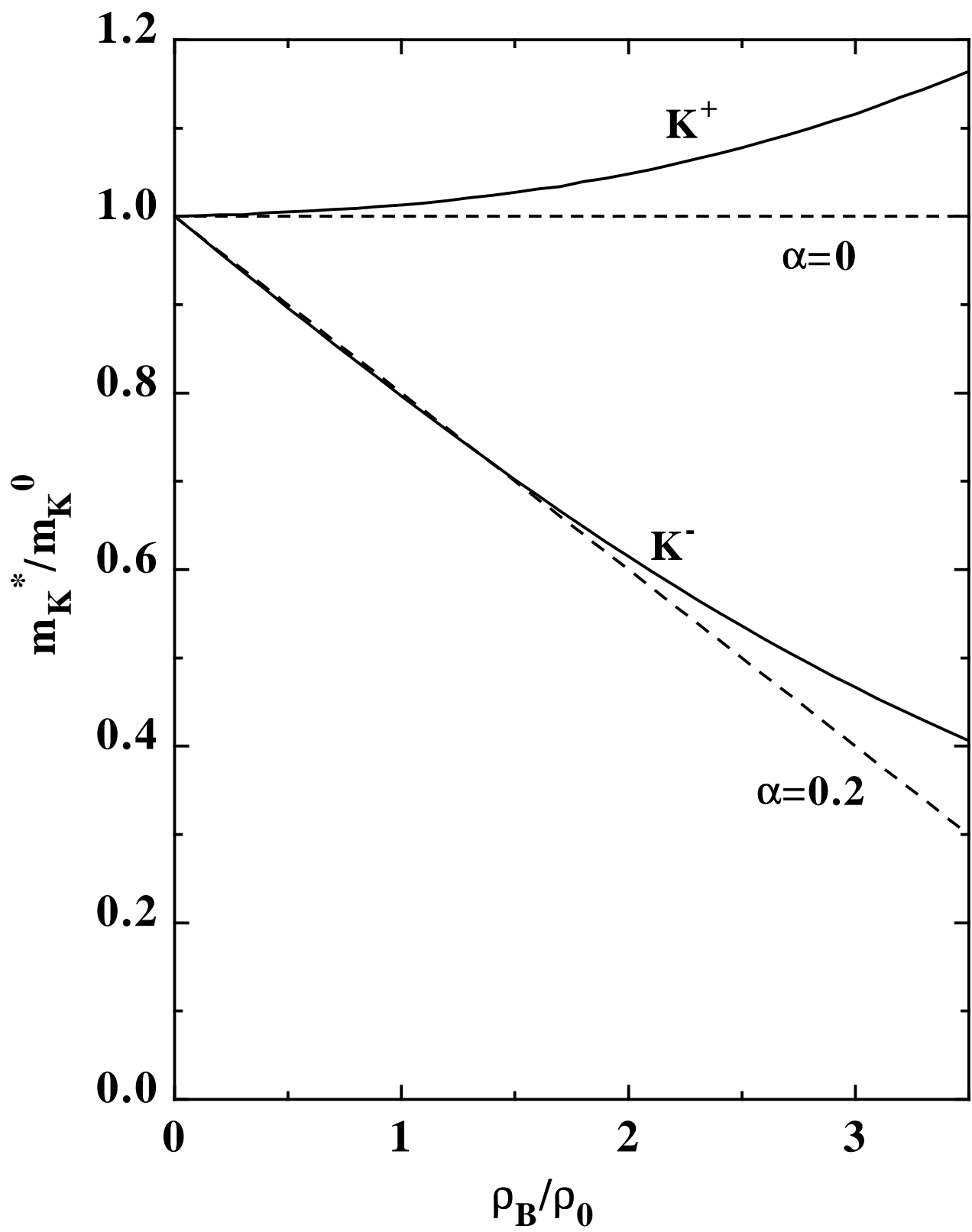
\begin{figure}[h]
\caption{
The $K^+$ and $K^-$ mass as a function of the baryon density in units
of $\rho_0 \approx 0.16 fm^{-3}$ according to Kaplan and Nelson
(Eq.~(\protect\ref{mg1})). The dashed line (with $\alpha$ = 0.2)
presents a linear fit to the $K^-$ effective mass.}
\label{Fig1}
\end{figure}

Before addressing the in-medium modifications of the antikaons, we show
our results for the inclusive $K^+$ invariant cross section for Ni~+~Ni
at 0.8, 1.0 and 1.8 GeV/u at $\theta_{lab} = $44$^o$ in
Fig.~\ref{Fig10}, that have been transformed to the nucleus-nucleus
cms. Also depicted in Fig.~\ref{Fig10} are the preliminary $K^+$
spectra for these systems from Ref.~\cite{Senger}. In comparison we
find that for positive kaons apparently no selfenergy effects are
needed to describe the data at the energies of interest in line with
the expected density dependence in Fig.~\ref{Fig1}. This finding is in
accordance with our earlier studies on $K^+$ production in
nucleus-nucleus~\cite{lang,Tomo,Cass90} and proton-nucleus
collisions~\cite{ca90a} in line with other groups~\cite{Hartnack}.

We now turn back to the system Ni~+~Ni at 1.85 GeV/u as well as 1.66
GeV/u and concentrate on the $K^-$ spectra at $0^o$ with respect to the
beam axis in the nucleus-nucleus cms. In Figs.~\ref{Fig11} and
\ref{Fig12} we show the effect of antikaon absorption on the spectra in
comparison with the data of Ref.~\cite{Schro} and the preliminary data
for Ni~+~Ni at 1.8 GeV/u from Ref.~\cite{Senger1}. The dashed lines
reflect calculations including the bare antikaon mass without any
antikaon absorption, while the dash-dotted lines include antikaon
absorption as in Fig.~\ref{Fig8}. We find that $K^-$ absorption reduces
the cross section on average by a factor of 5 for the Ni~+~Ni systems.

We now address the possible medium effects on the antikaon according to
Fig.~\ref{Fig1} or Eq.~(\ref{kmass}), respectively. We note that for
$\alpha$ = 0 we recover the limit of vanishing antikaon selfenergy,
whereas for $\alpha \approx$ 0.2  we describe the scenario of Kaplan
and Nelson~\cite{Kaplan}\footnote{For practical purposes one should
consider $\alpha $ to be a free parameter to be fixed in comparison to
the experimental data in order to learn about the magnitude of the
antikaon selfenergy.  In fact, we obtain a much better reproduction of
the spectra at both bombarding energies for $\alpha \approx 0.24$, but
due to the uncertainties involved in the elementary $BB$ production cross
sections we cannot determine this value very reliably.}.  In
Figs.~\ref{Fig11} and \ref{Fig12} we also display the results of
calculations for $\alpha$ = 0.2 (solid lines).  With increasing
$\alpha$ not only the magnitude of the spectra is increased, but also
the slope becomes softer. For $\alpha \approx 0.2$ (in line with
Fig.~\ref{Fig1}) we still underestimate the experimental spectra
slightly. In terms of Eq.~(\ref{mg1}) this would imply that the
'fundamental' quantity $\Sigma _{KN}$ might be even larger than 350
MeV. Anyway, by assuming quite sizeable antikaon attractive
selfenergies we can reproduce the data.

\section{Summary}
In this work we have presented a detailed study of kaon and antikaon
production in Ni~+~Ni collisions from 0.8 to 1.85 GeV/u within the
coupled channel BUU approach, where the kaons and antikaons are
produced perturbatively, however, propagated explicitly with their
final state interactions. An important ingredient of our reanalysis of
the kaon production cross sections in heavy-ion collisions are the
novel elementary production cross sections from
Refs.~\cite{Sibirt1,Sibirt2,Sibirt3,Sibirt4} (cf. Sect. 2.2), that are
much lower than previous parametrizations used so far in the
literature~\cite{zwermann,randrup}. Furthermore, we have incorporated
for the first time all relevant production channels for kaons that are
known or can be deduced from hadronic reactions in free space.

Our analysis shows that $\pi^-$ and $K^+$ spectra are reasonably  well
described in this energy regime without introducing any medium
modifications for these mesons (cf. Ref.~\cite{Teis96} in case of
pions).  This experience is fully in line with our earlier studies on
this subject and the results from independent
groups~\cite{Hartnack,Bass}.  The antikaon spectra, however, are
underestimated severely when incorporating only bare kaon masses
roughly in line with the study by Li et al.~\cite{Kolix}. These
authors, however, have used extrapolations for the antikaon cross
section, which -- to our present knowledge -- are severely
overestimated (cf. Fig.~\ref{Fig2}).  When including an attractive
antikaon potential comparable to that proposed by Kaplan and
Nelson~\cite{Kaplan}, a satisfactory description of the $K^-$ spectra
can be given, both in the actual magnitude as well as in the slope. It
is worth noting that the $\pi$-hyperon production channels play a
sizeable role in case of the vacuum antikaon mass (Fig.~\ref{Fig8}),
whereas their contribution  in the 'dropping mass scenario' of Kaplan
and Nelson~\cite{Kaplan} becomes of minor importance.

The dropping of the antikaon mass with baryon density may be
interpreted as a step towards a partial restoration of chiral symmetry
that can already be observed at SIS energies. Similar observations have
been made at AGS energies~\cite{Ehehalt} as well as at SPS energies,
where especially a dropping of the $\rho$-mass can be used to
accurately describe the dilepton spectra from heavy-ion
reactions~\cite{Ca95,LKB,Ca96}. Though there are quite a number of
indications for dropping meson masses in the medium by now, one has to
properly examine the possibility that conventional many-body effects
such as $\Lambda$-hole loops~\cite{wass} or 'resonance-hole'
loops~\cite{friman} may also account for the spectra observed. These
could also affect the selfenergies of hyperons, which are presently
treated as 'heavy nucleons'. In addition, we mention again that
experimental studies of antikaon production in NN reactions close to
threshold are urgently needed to have a final control on the input
cross sections used in our present analysis.  Another uncertainty in
our analysis, i.e. the behaviour of the $\Delta N$ production cross
sections, can be eliminated by performing studies for $K^+$ and $K^-$
production in proton-nucleus reactions, where the $\Delta$-excitation
plays a minor role~\cite{Demski}.

\vspace{1.5cm}
\noindent
The authors acknowledge valuable and inspiring discussions throughout
this work with C.~M.~Ko, P.~Senger and Gy.~Wolf.

\newpage
\section*{Figure Captions}

\begin{figure}[h]
\caption{
The parametrizations for the isospin averaged inclusive $K^+$ and $K^-$
cross sections from $NN$ collisions as a function of the invariant
energy $\protect\sqrt{s}$ above threshold $\protect\sqrt{s_0}$ (solid
lines) in comparison to the experimental data \protect\cite{Giacom}.
The dash-dotted line is the parametrization from
Ref.~\protect\cite{zwermann} for $K^-$ production whereas the open
square (denoted by OBE) is from the OBE calculation of
Ref.~\protect\cite{Sibirt4}.  }
\label{Fig2}

\vspace*{5mm}
\caption{
Comparison of our parametrizations with the experimental cross sections
\protect\cite{landolt} for the isospin channels $pp \to pnK^+\bar{K}^0,
pp \to p\Lambda K^+, pp \to p\Sigma^0 K^+, pp \to p \Sigma^+ K^0$.  }
\label{Fig3}

\vspace*{5mm}
\caption{
$K^-N$ inelastic (solid line) and elastic (dashed line) cross section
as a function of kaon laboratory momentum $p_K$ as fitted to the
experimental data from \protect\cite{landolt}.  The dotted line
displays the elastic $K^+N$  cross section.  }
\label{Fig4}

\vspace*{5mm}
\caption{
Antikaon production cross section from $\pi$ + hyperon collisions
according to Eq.~(\protect\ref{Sppkp1}). The dash-dotted line
represents the channel $\pi \pi \to K\bar{K}$ according to
Eq.~(\protect\ref{pipi}).  }
\label{Fig5}

\vspace*{5mm}
\caption{
Antikaon production cross sections from hyperon-nucleon collisions
according to an OBE calculation.  }
\label{Fig6}

\vspace*{5mm}
\caption{
The inclusive Lorentz-invariant cross section as a function of the
meson momentum  in the nucleus-nucleus cms for $\pi^-$ and $K^-$ mesons
at $\theta_{lab} = 0^o$ for Ni~+~Ni at 1.85 GeV/u.  The full dots and
full squares represent the experimental data from
Ref.~\protect\cite{Schro}.  In these calculations no meson selfenergies
have been taken into account.  }
\label{Fig7}
\end{figure}

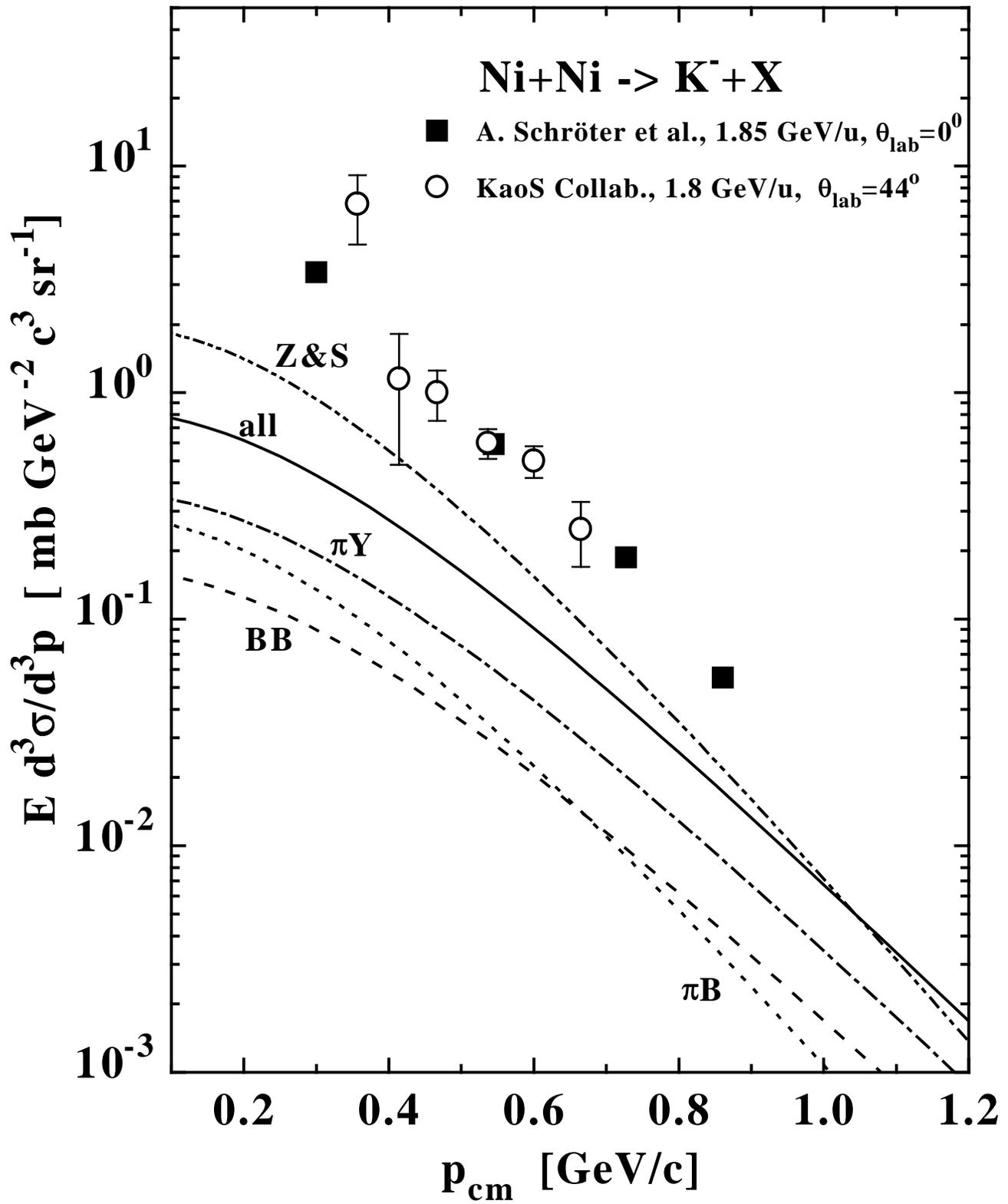
\begin{figure}[h]
\caption{
The inclusive Lorentz-invariant cross section as a function of the kaon
momentum  in the nucleus-nucleus cms for $K^-$ mesons at $\theta_{lab}
= 0^o$ for Ni~+~Ni at 1.85 GeV/u without including $K^-$ selfenergies
in comparison to the experimental data from \protect\cite{Schro} and the
preliminary data for Ni~+~Ni at 1.8 GeV/u from \protect\cite{Senger1}.
The dashed line displays the cross section from baryon-baryon ($BB$)
channels, the dotted line that from pion-baryon ($\pi B$) channels,
while the dash-dotted line shows the contribution from $\pi$-hyperon
collisions.  The solid line represents the sum of all channels taken
into account. In addition we show the full result when using the
parametrization from Ref.~\protect\cite{zwermann} for the $BB$ channel. }
\label{Fig8}
\end{figure}

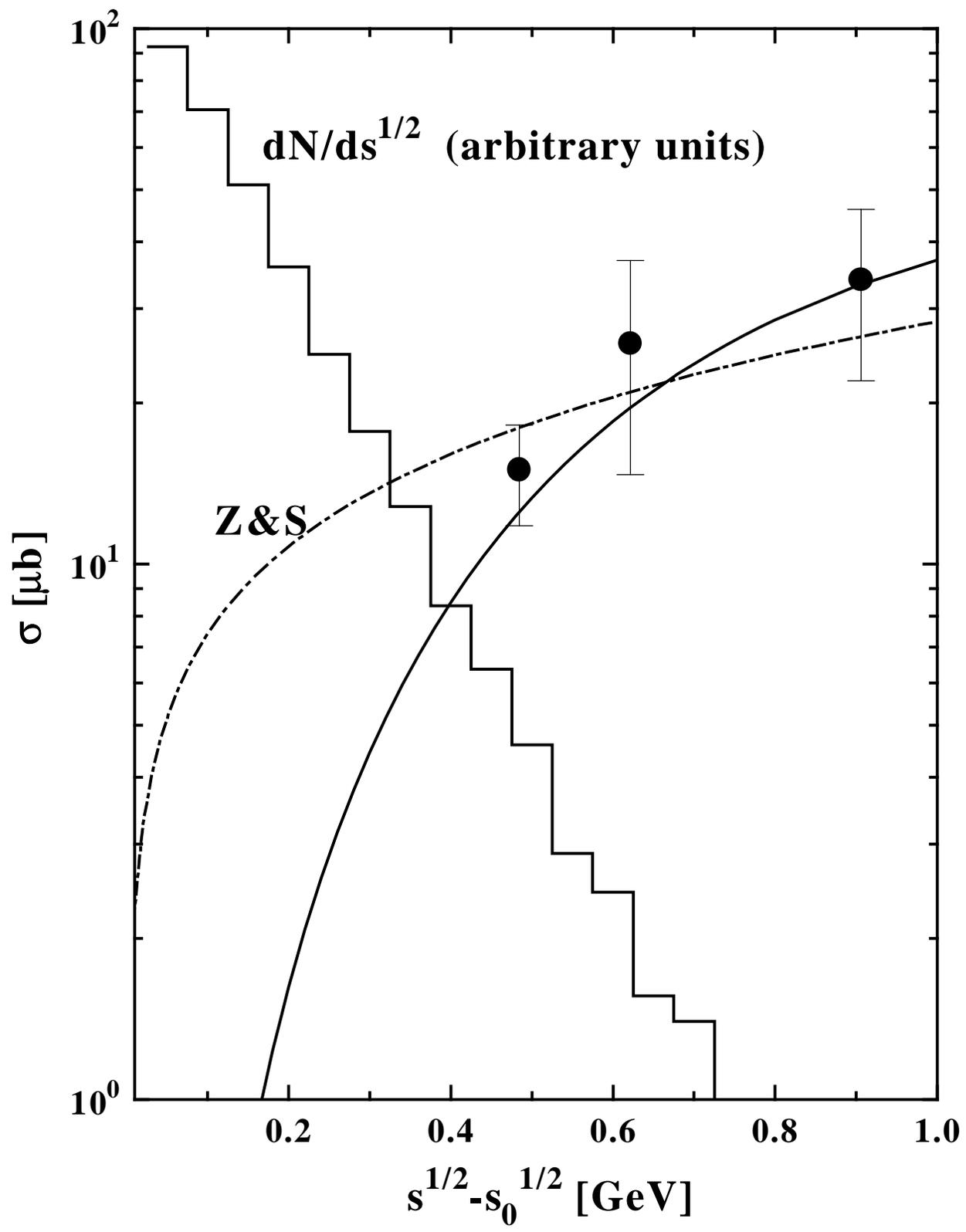
\begin{figure}[t]
\caption{
The distribution in the $BB$ collision number versus the invariant
energy $dN/ds^{1/2}$ for Ni~+~Ni at 1.85 GeV/u (histogram).  For
orientation we display again our parametrization for the isospin
averaged inclusive $K^-$ cross section (solid line) in comparison to
the experimental data and the parametrization from
Ref.~\protect\cite{zwermann} (dot-dashed line).  }
\label{Fig9}
\end{figure}

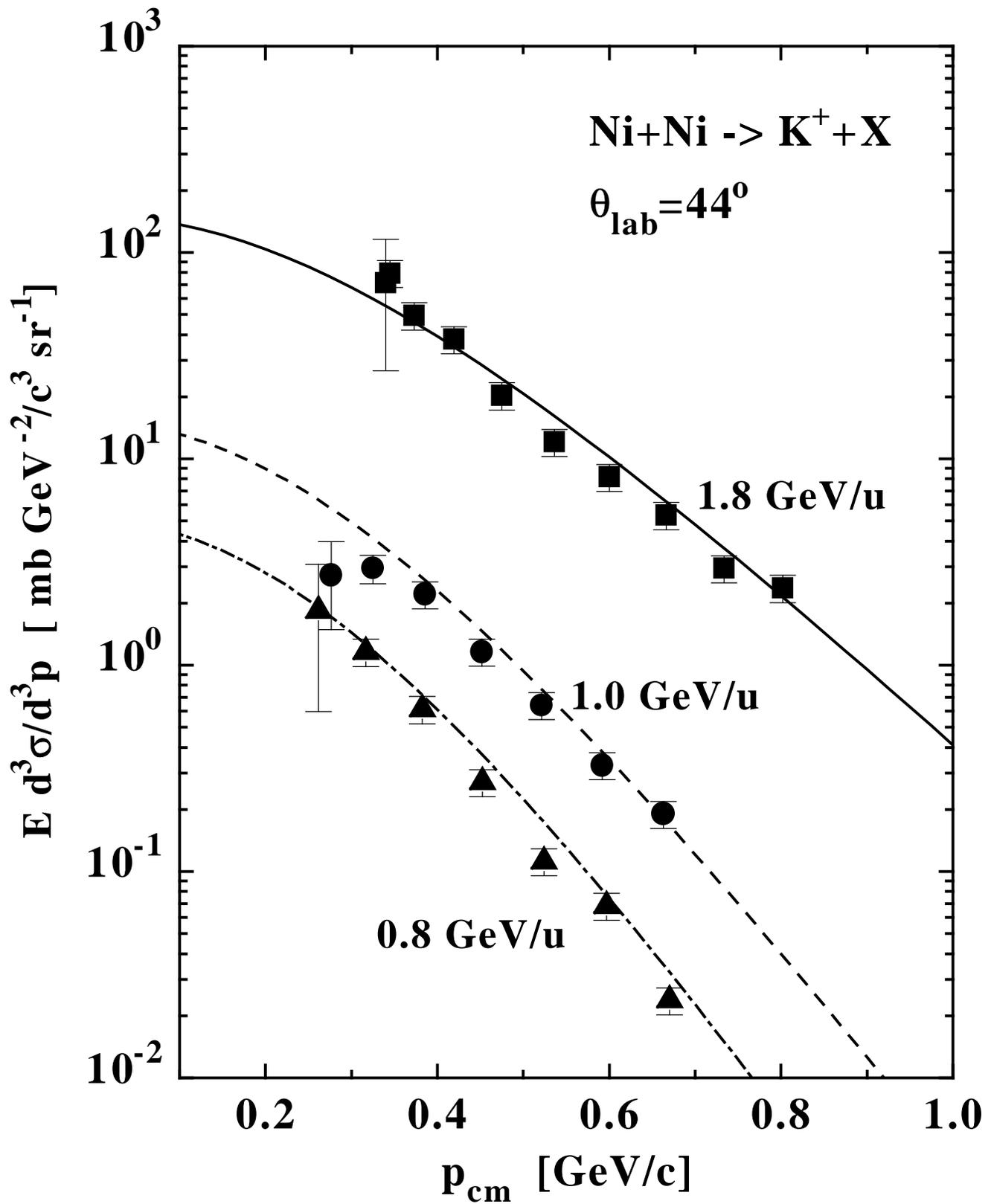
\begin{figure}[h]
\caption{
The inclusive Lorentz-invariant cross section as a function of the
meson momentum  in the nucleus-nucleus cms for $K^+$ mesons at
$\theta_{lab} = 44^o$ for Ni~+~Ni at 0.8, 1.0 and 1.8 GeV/u without
including any meson selfenergies. The experimental preliminary $K^+$
spectra have been taken from Ref.~\protect\cite{Senger}.  }
\label{Fig10}
\end{figure}

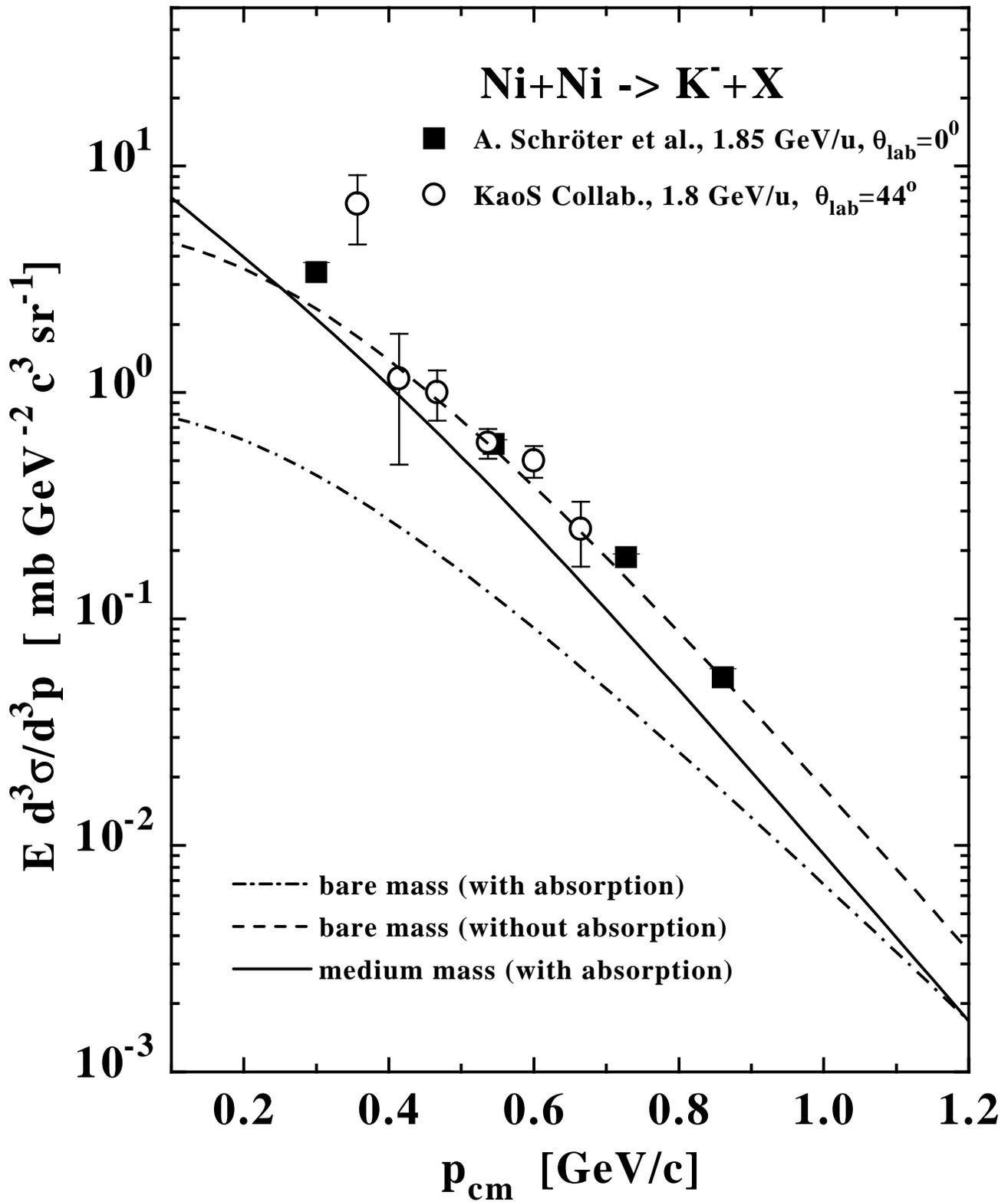
\begin{figure}[h]
\caption{
The inclusive Lorentz-invariant cross section as a function of the kaon
momentum  in the nucleus-nucleus cms for  $K^-$ mesons at $\theta_{lab}
= 0^o$ for Ni~+~Ni at 1.85 GeV/u in comparison to  the experimental
data from Ref.~\protect\cite{Schro} and the preliminary data for Ni~+~Ni
at 1.8 GeV/u from Ref.~\protect\cite{Senger1}.  The dashed line
corresponds to a calculation with bare kaon masses when discarding
$K^-$ reabsorption, whereas the dash-dotted line includes $K^-$
reabsorption. The solid line displays the results with antikaon
selfenergies Eq.~(\protect\ref{kmass}) for $\alpha$ = 0.2 including
$K^-$ reabsorption.  }
\label{Fig11}
\end{figure}

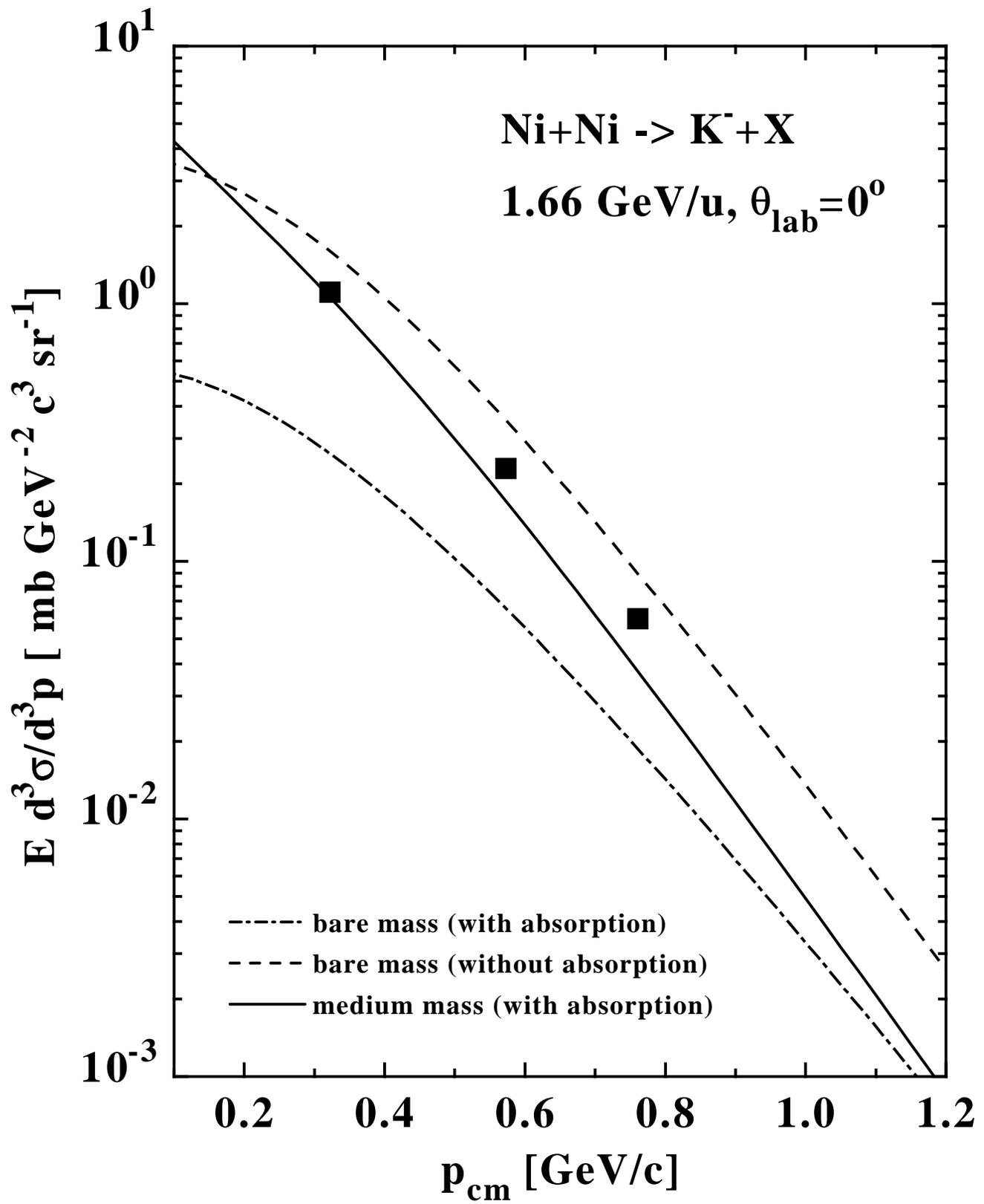
\begin{figure}[h]
\caption{
The inclusive Lorentz-invariant cross section as a function of the kaon
momentum  in the nucleus-nucleus cms for  $K^-$ mesons at $\theta_{lab}
= 0^o$ for Ni~+~Ni at 1.66 GeV/u in comparison to  the experimental
data from Ref.~\protect\cite{Schro}.  The dashed line corresponds to a
calculation with bare kaon masses when discarding $K^-$ reabsorption,
whereas the dash-dotted line includes $K^-$ reabsorption. The solid
line displays the results with antikaon selfenergies
Eq.~(\protect\ref{kmass}) for $\alpha$ = 0.2 including $K^-$
reabsorption.  }
\label{Fig12}
\end{figure}

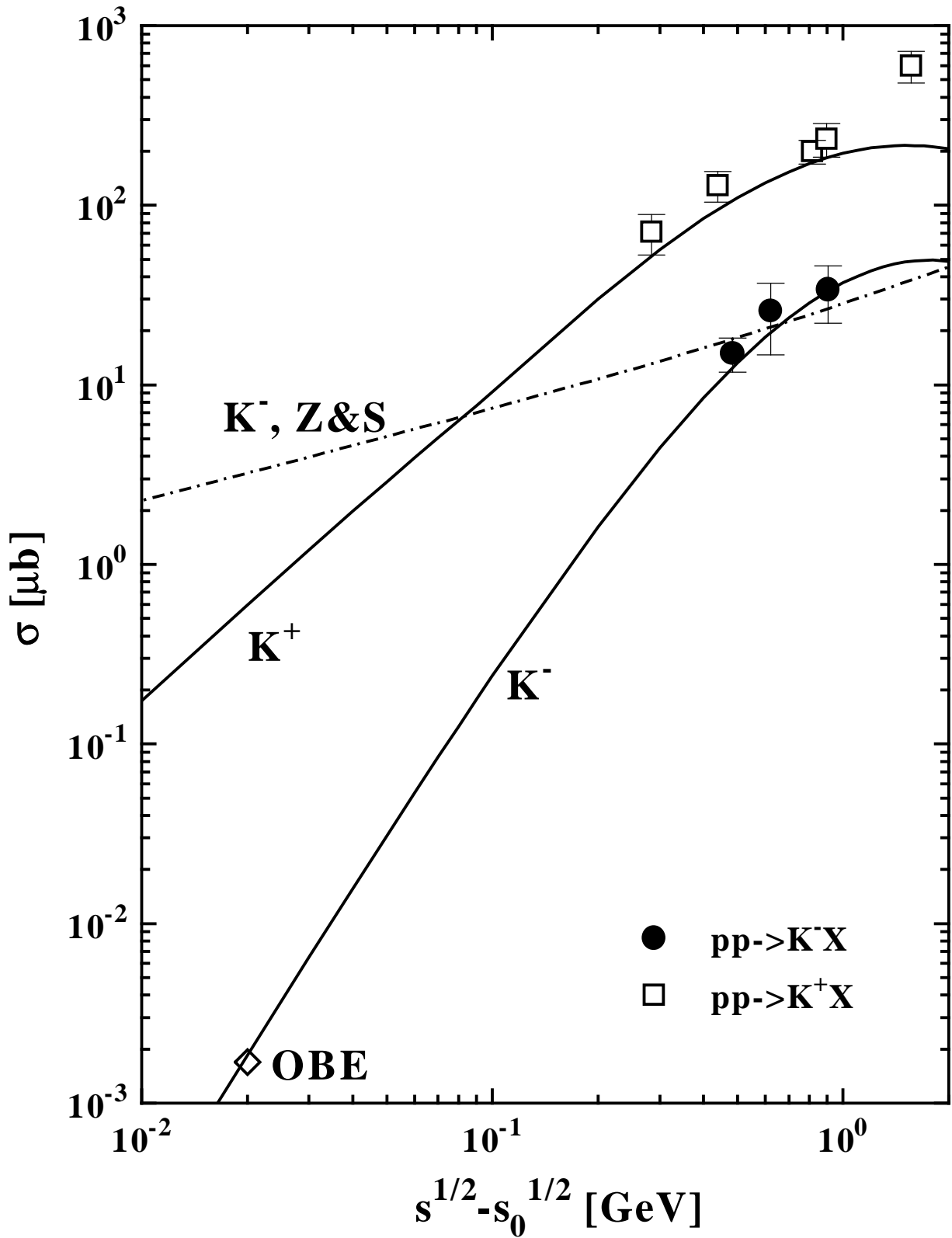
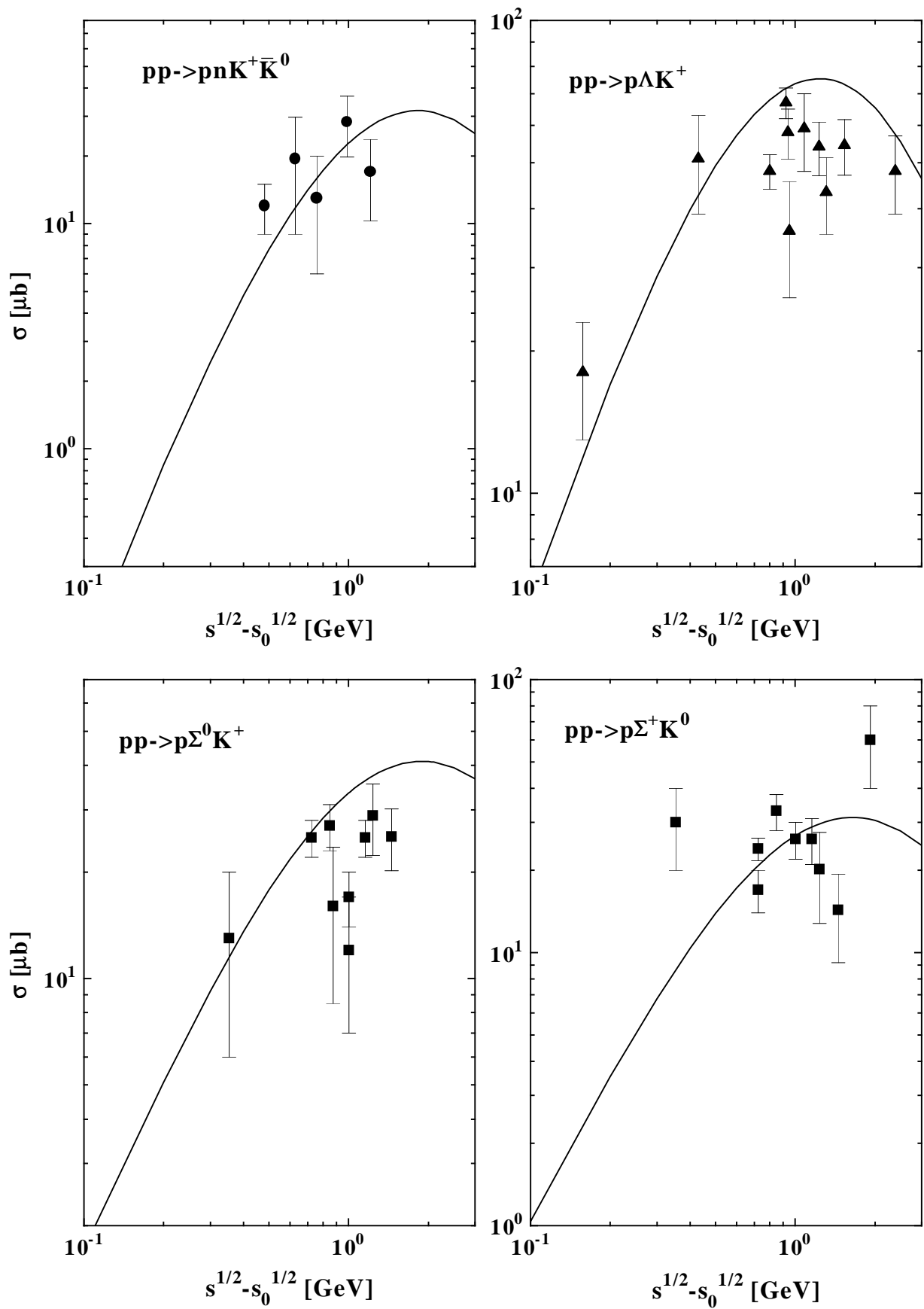
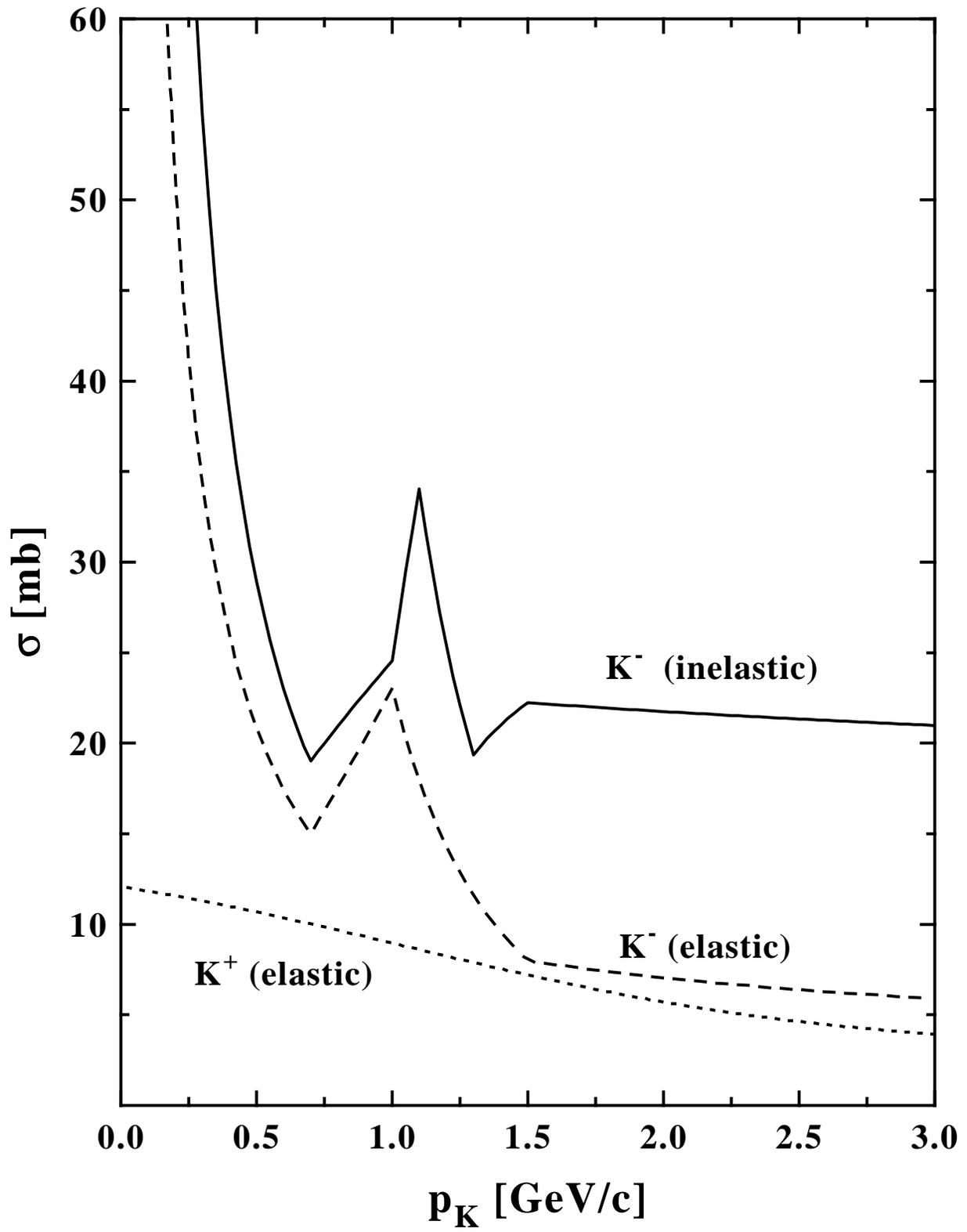
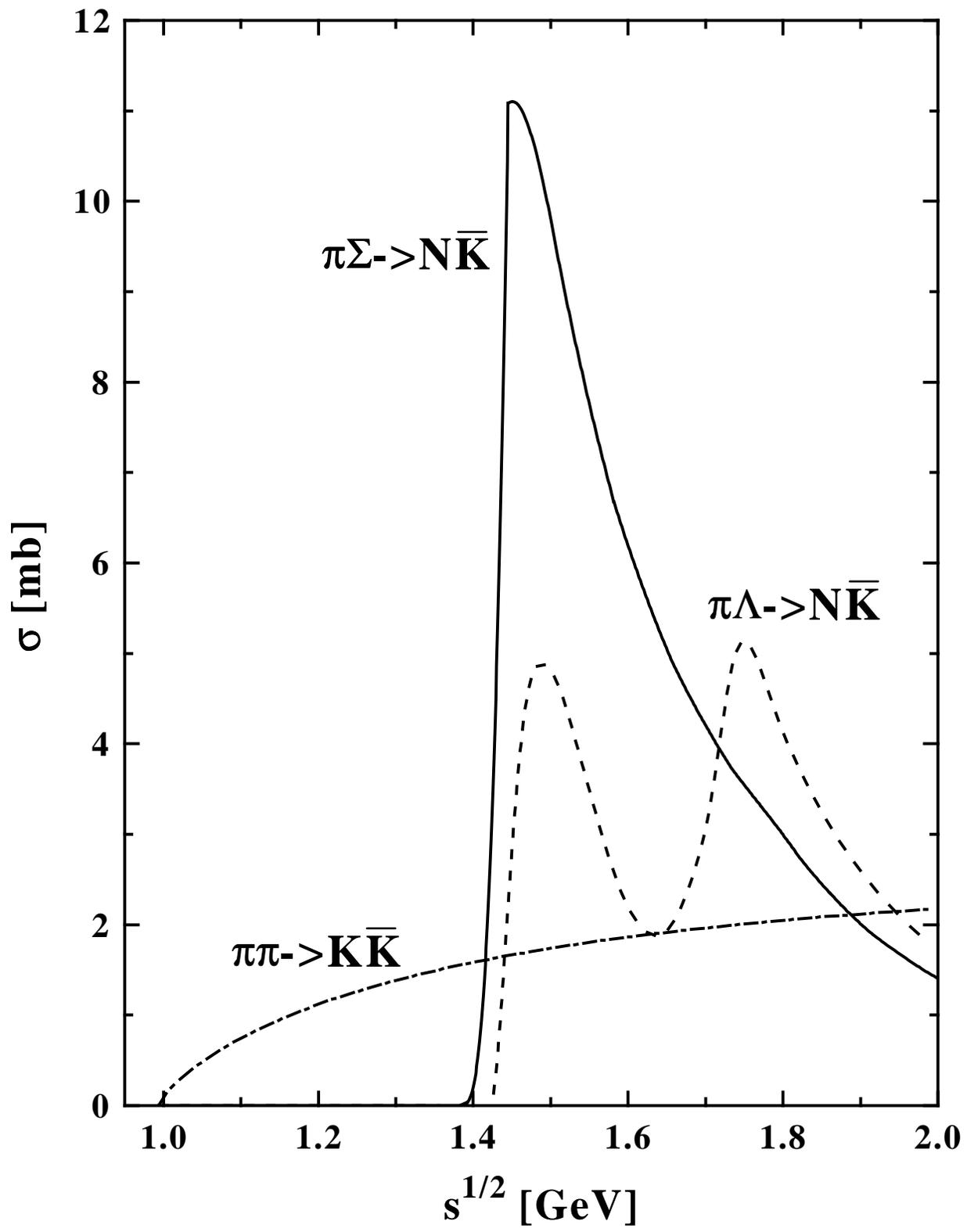
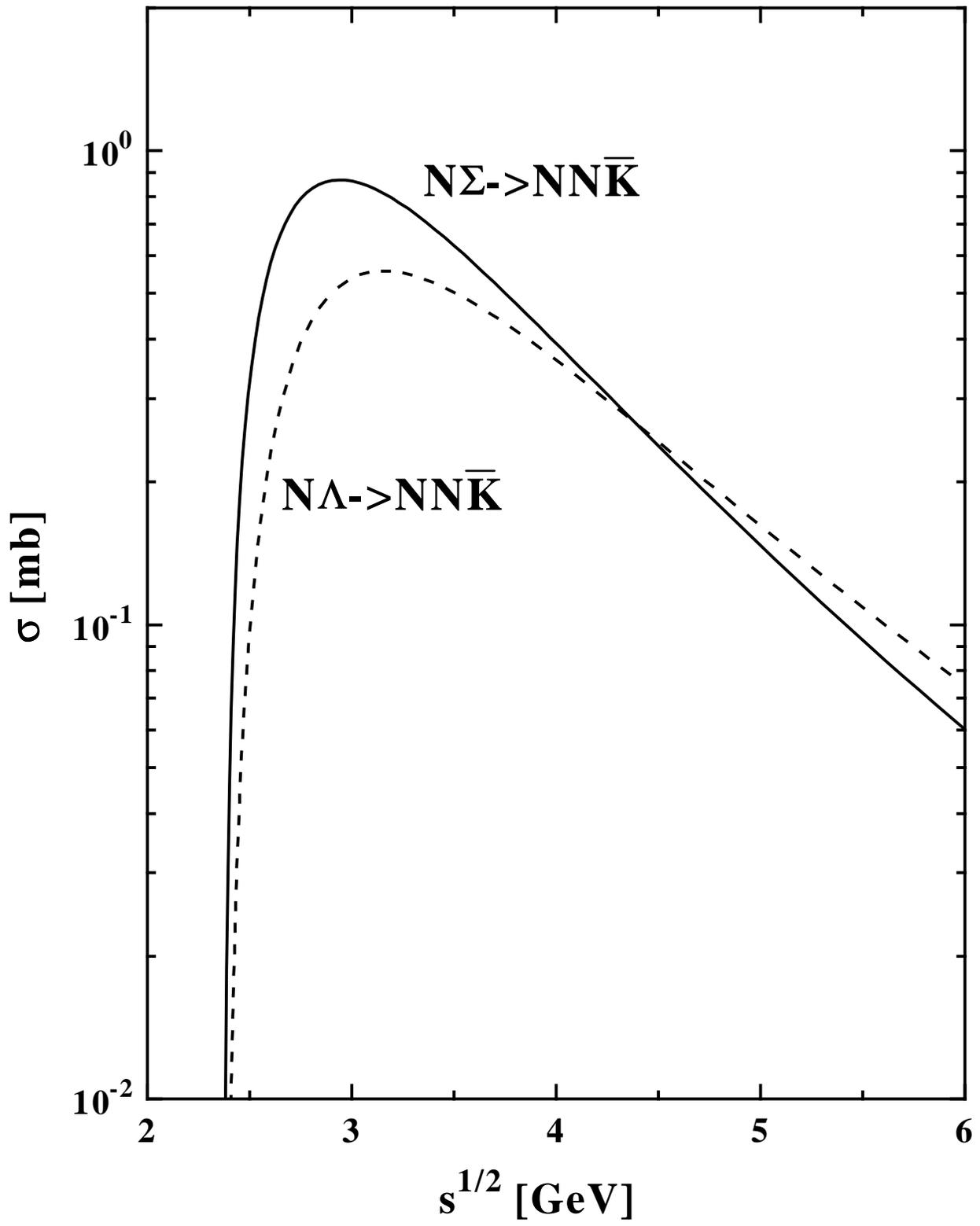
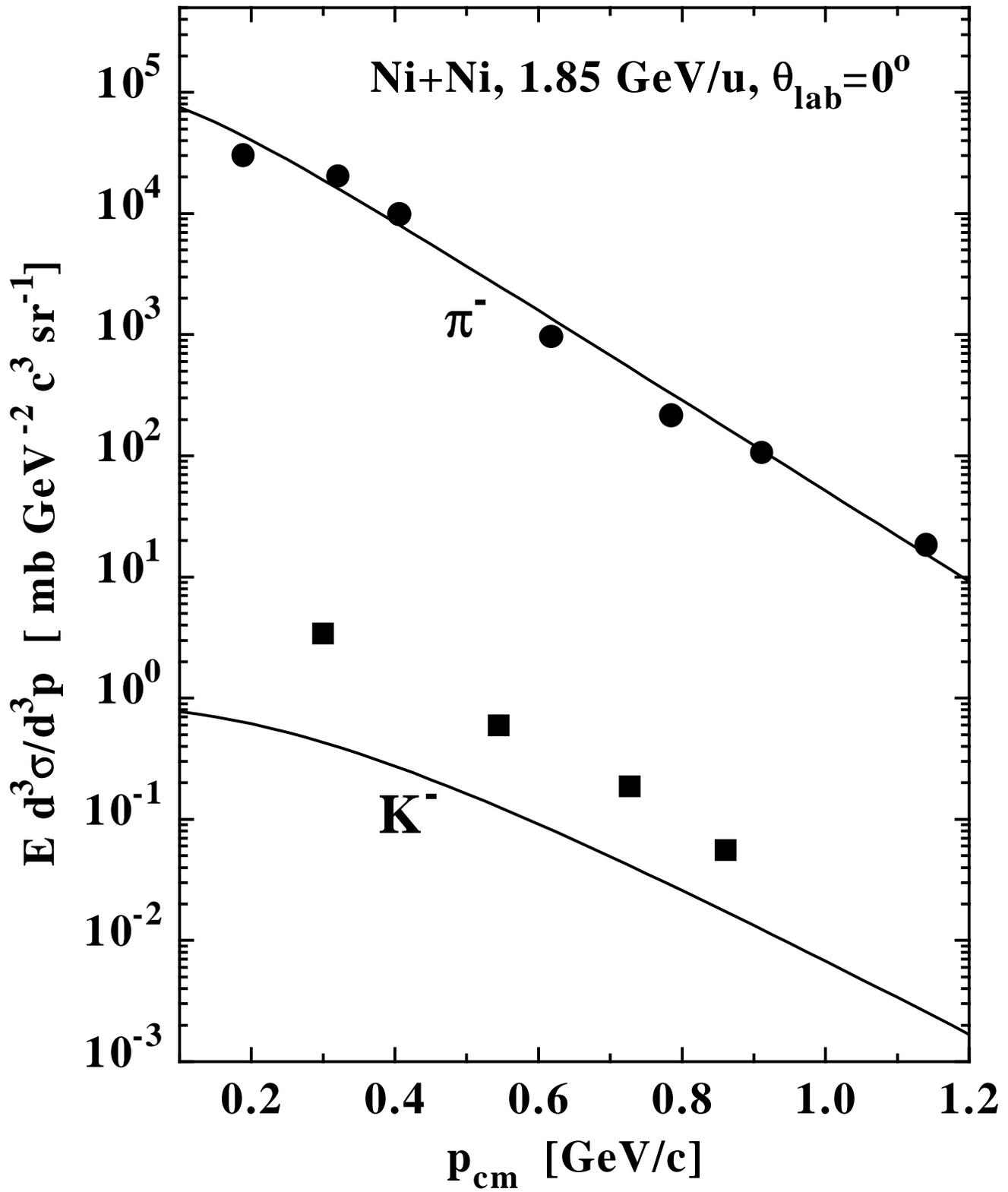
\end{document}